\documentclass[preprint]{aastex62}
\usepackage{subfigure}
\usepackage{graphicx}
\usepackage{epsfig}

\newcommand{\bjdtdb}{\ensuremath{\rm {BJD_{TDB}}}}
\newcommand{\feh}{\ensuremath{\left[{\rm Fe}/{\rm H}\right]}}
\newcommand{\teff}{\ensuremath{T_{\rm eff}}}

\newcommand{\msun}{\ensuremath{\,M_\Sun}}
\newcommand{\rsun}{\ensuremath{\,R_\Sun}}

\received{April 8, 2018}
\accepted{May 9, 2018; Accepted in The Astronomical Journal}

\shorttitle{EPIC~211945201}
\shortauthors{Chakraborty et al.}

\begin{document}

\title{EVIDENCE OF A SUB-SATURN AROUND EPIC~211945201}

\correspondingauthor{Abhijit Chakraborty}
\email{abhijit@prl.res.in}

\author{Abhijit Chakraborty}
\affil{Astronomy \& Astrophysics Division, Physical Research Laboratory, Ahmedabad 380009, India}

\author[0000-0001-8127-5775]{Arpita Roy}
\affil{Department of Astronomy, California Institute of Technology, Pasadena, CA 91125, USA\\}

\author{Rishikesh Sharma }
\affiliation{Astronomy \& Astrophysics Division, Physical Research Laboratory, Ahmedabad 380009, India}

\author[0000-0001-9596-7983]{Suvrath Mahadevan}
\affil{Department of Astronomy \& Astrophysics, The Pennsylvania State University, University Park, PA 16802, USA}
\affiliation{Center for Exoplanets and Habitable Worlds, The Pennsylvania State University, University Park, PA 16802, USA \\}

\author{Priyanka Chaturvedi}
\affiliation{Astronomy \& Astrophysics Division, Physical Research Laboratory, Ahmedabad 380009, India}

\author{Neelam J.S.S.V Prasad}
\affiliation{Astronomy \& Astrophysics Division, Physical Research Laboratory, Ahmedabad 380009, India}

\author{B. G. Anandarao}
\affiliation{Astronomy \& Astrophysics Division, Physical Research Laboratory, Ahmedabad 380009, India}

\begin{abstract}
We report here strong evidence for a sub-Saturn around EPIC~211945201 and confirm its planetary nature. EPIC~211945201b was found to be a planetary candidate from {\it K2} photometry in Campaigns 5 \& 16, transiting a bright star ($V_{\rm mag}=10.15$, G0 spectral type) in a 19.492 day orbit. However, the photometric data combined with false positive probability calculations using VESPA was not sufficient to confirm the planetary scenario. Here we present high-resolution spectroscopic follow-up of the target using the PARAS spectrograph (19 radial velocity observations) over a time-baseline of 420 days. We conclusively rule out the possibility of an eclipsing binary system and confirm the 2-$\sigma$ detection of a sub-Saturn planet. The confirmed planet has a radius of 6.12$\pm0.1$$~R_{\oplus}$, and a mass of $27_{-12.6}^{+14}$~$M_{\oplus}$. We also place an upper limit on the mass (within the 3-$\sigma$ confidence interval) at 42~$M_{\oplus}$ above the nominal value. This results in the Saturn-like density of $0.65_{-0.30}^{+0.34}$ g~cm$^{-3}$. Based on the mass and radius, we provide a preliminary model-dependent estimate that the heavy element content is 60-70 \% of the total mass. This detection is important as it adds to a sparse catalog of confirmed exoplanets with masses between 10-70 $M_{\earth}$ and radii between 4-8 $R_{\earth}$, whose masses and radii are measured to a precision of 50\% or better (only 23 including this work).

\end{abstract}


\section{Introduction} \label{sec:intro}

A large number of transiting exoplanets have been discovered by dedicated space based photometric missions such as {\it CoRoT} \citep{Baglin2006}, {\it Kepler} \citep{Borucki2010} and {\it K2} \citep{Howell2014}, and then followed up with ground-based spectroscopic resources for mass measurement using the radial velocity (RV) technique. However, the canon is limited in the number of exoplanets with radii between 2-8 $R_{\earth}$ that also have masses measured with a precision of 50\% or better. Subsequently, this has limited our understanding of the composition, evolutionary history, and diversity of a population of exoplanets that have been variously defined as super-Neptunes \citep[10-40 $M_{\earth}$ and 2-6 $R_{\earth}$,][]{Barragan2016} or sub-Saturns \citep[10-70 $M_{\earth}$ and 4-8 $R_{\earth}$,][]{Petigura2017}.

In the absence of precise mass measurements, or rather prior to the engagement of ground-based spectroscopic resources, there are still confidence levels that can be placed on the detection of an exoplanet candidate. Statistical validation tools such as BLENDER \citep{Torres2011}, PASTIS \citep{Diaz2014}, and VESPA \citep{Morton2012}, can give higher confidence in the planetary scenerio than false positive alternatives by calculating the authenticity of the transit signal, in conjunction with ancillary information about the system, using a Bayesian approach. For example, recent work by \citet{Mayo2018} confirmed 149 exoplanets and identified 275 planetary candidates (PC) based on the false positive probabilities (FPP) of the transit signature. However, there remain sources whose transit signals are inadequate to confirm the nature of the system -- these are essential candidates for follow-up with high-precision RV spectrographs. Combining the photometry with RV data allows us to determine the mass and radius of the exoplanet, and hence its density. Theories of internal structure and planet formation mechanism are increasingly better constrained as we expand the number of exoplanets with precise measurements of mass and radius \citep[and references therein]{Lopez&Fortney2014}. Hence ground-based radial velocity follow-up for mass determination, although quite resource-limited, remains of extreme importance for understanding exoplanet demographics.

In this paper, we report evidence for the sub-Saturn nature of EPIC~211945201b, a planetary candidate observed in {\it K2} Campaigns 5 \& 16. In Sections~\ref{sec:photometry} \& \ref{sec:imaging} we present the {\it K2} photometry and the analysis of archival Keck $K$-band imaging data. Section~\ref{sec:vespa} elaborates on the statistical validation procedures of the PC using the VESPA framework. In Section~\ref{sec:rv_obs} we describe our follow-up campaign with the PARAS spectrograph and the corresponding RV analysis. Section~\ref{sec:isochrones} reports our final host star properties, and Section~\ref{sec:pyaneti} describes the simultaneous fitting of RV and photometric data. We discuss our results in Section~\ref{sec:internal_comp} and conclude in Section~\ref{sec:summary}.

\section{{\it K2}-photometry} \label{sec:photometry}

The NASA€™ {\it K2} mission \citep{Howell2014} observed the source EPIC~211945201 from 27 April 2015 to 10 July 2015 (73 days) \& 07 December 2017 to 25 February 2018 (81 days) as a part of Campaign 5 and Campaign 16, respectively. This target was identified as a transiting planetary candidate system by \citet{Pope2016}, \citet{Barros2016}, \citet{Petigura2018} and \citet{Mayo2018} using the light curve from Campaign 5, while \citet{Yu2018} recently used campaign 16 data to declare it a high quality planetary candidate.

There have been several pipelines developed to correct the systematics from the {\it K2} light curve. Examples include {\tt{k2sff}}  \citep{Vanderburg2014}, {\tt{k2sc}} \citep{Aigrain2016}, {\tt{k2phot}} \citep{Aigrain2016} \& the {\tt{everest}} package \citep{Luger2016}. All the works identifying the possible planetary nature of EPIC~211945201b, referred to above, used one of above pipelines to extract and correct the light curve. In general, they first corrected the light curve and then searched for the significant transit signature using the {\tt{BLS}} \citep{Kovacs2002} algorithm. In the BLS algorithm, the transit event is modeled as a box shaped modulation of the light curve. Following this approach, the occurrence of a single transit event was found at 19.49179 days \citep{Mayo2018} from Campaign 5, and at 19.492036 days \citep{Yu2018} from Campaign 16, which are both consistent within their error bars.

\begin{figure*}[!ht]
	\includegraphics[width=1.0\textwidth]{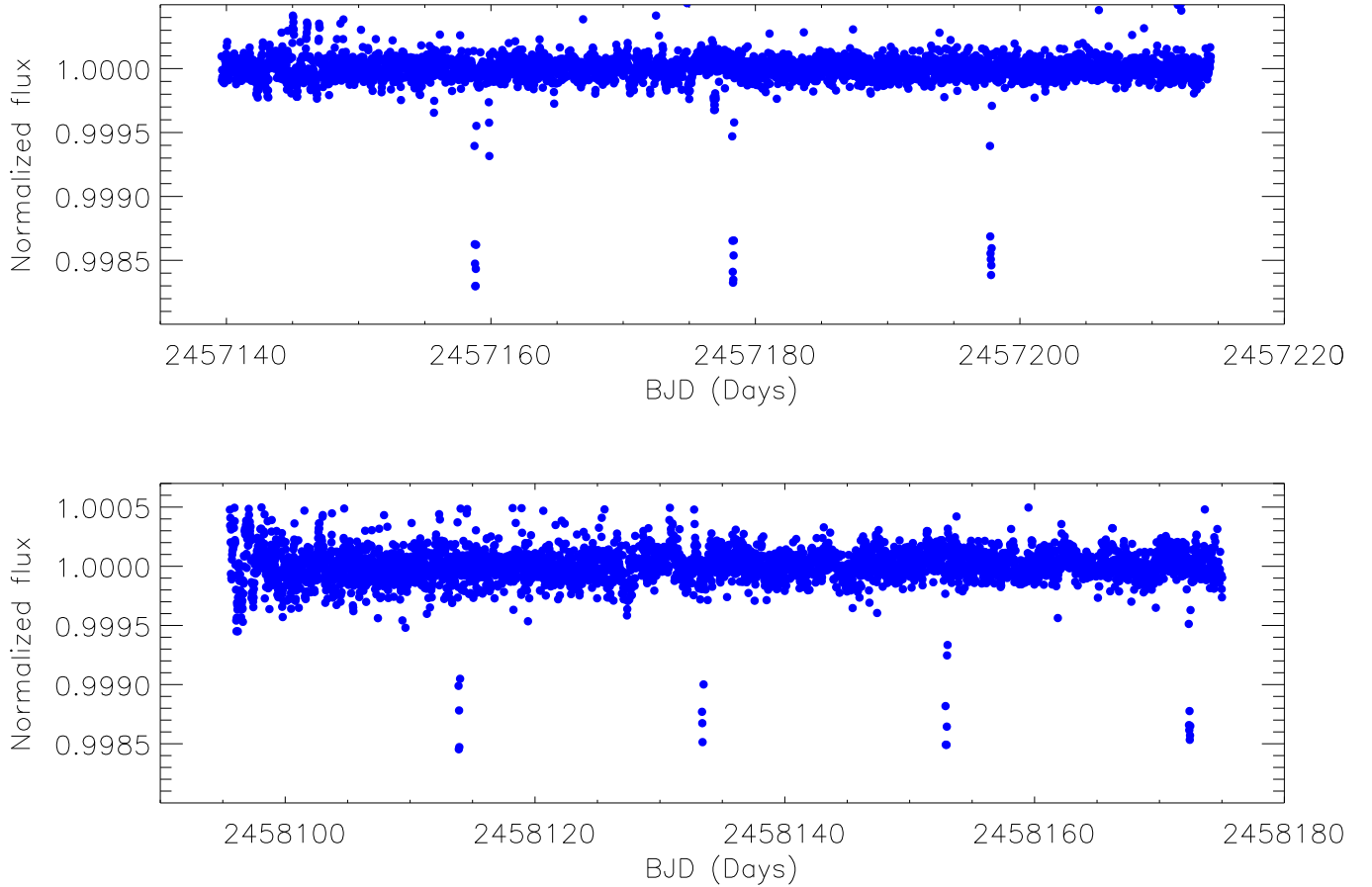}
\caption{{\tt K2PHOT} photometry data of EPIC~211945201. The upper panel shows the light curve from Campaign 5 ($\approx{73}$ d), while the lower panel light curve is from Campaign 16 ($\approx{81}$ d). Three transits in Campaign 5, and four transits in Campaign 16 are clearly visible for this target. \label{fig:light_curve}}
\end{figure*}

We retrieved the {\tt K2PHOT} light curves \citep{Petigura2015, Aigrain2016} for both campaigns through ExoFOP\footnote{https://exofop.ipac.caltech.edu/k2/} -- these are shown in Fig~\ref{fig:light_curve}.  Seven transits, three in the upper panel \& four in the lower panel, spaced every $\sim 19.49$ days, are clearly visible in the light curve of EPIC~211945201. In this work we use the combined light curve of both Campaigns 5 \& 16, and adopt a period of 19.49215 days from its analysis (see \S\ref{sec:pyaneti}).

\section{High Angular Resolution Imaging}
\label{sec:imaging}

\begin{figure*}[!ht]
\centering
	\includegraphics[width=0.5\textwidth]{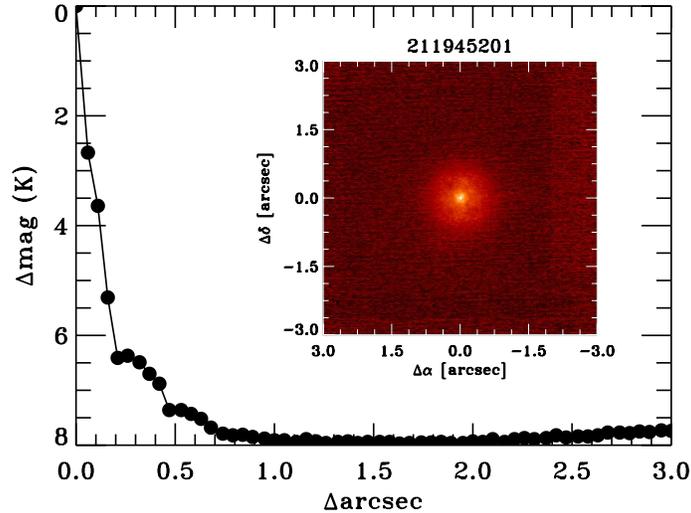}
\caption{High angular resolution image of the source EPIC~211945201, and the 5-$\sigma$ sensitivity/contrast curve curve in $K$-band observed with the NIRC2 instrument on the 10m Keck II telescope (publicly available on ExoFOP, contributed by user David Ciardi). \label{fig:contrast_curves}}
\end{figure*}

High-resolution imaging is an immensely useful tool for constraining the probability of blended background objects. We made use of the archival high angular resolution AO imaging data acquired using the Keck II/NIRC2 camera. This is also publicly available on ExoFOP (uploaded by user David Ciardi). The Keck AO data was obtained on January 21st of 2016(UT) with a $K$-band filter. NIRC2 has a pixel scale of 0.009942\arcsec$/$pixel. The observed image and 5-$\sigma$ sensitivity curve are shown in Fig~\ref{fig:contrast_curves}. The star appears single and has no close companions within the several arcseconds. At a separation of 0.5\arcsec the estimated sensitivity to the companion is $\approx{8}$ mag. This effectively rules out the possibility of background sources within this separation contributing significant flux to the light curve. The estimated point spread function (PSF) of the source is 0.0526\arcsec.

\section{Statistical Validation} \label{sec:vespa}

We used the open source and publicly available {\tt VESPA}\footnote{https://github.com/timothydmorton/VESPA} package from \citet{Morton2012,Morton2015}, to determine the FPP (False Positive Probabilities) of the transit signature. The code uses transit parameters like shape, depth, and duration of each transit event, as well as independent observational constraints (like AO imaging for example) to validate the transit signal of a planet. {\tt VESPA}
uses TRILEGAL (TRIdimensional modeL of thE GALaxy) to simulate the population of each possible false positive  scenario (such as background eclipsing binary, eclipsing binary, hierarchical eclipsing binary etc.) in a particular part of the sky using the coordinates of the source. It uses this population set combined with the observational constraints/priors to calculate the prior for each scenario, and then the likelihood of the each scenario. The resultant numbers are used to calculate a final probability for the validation of each contamination scenario.

{\tt VESPA} requires certain inputs before calculating the FPP, such as aperture size used for the extraction of the light curve, secondary eclipse threshold, and photometric and spectroscopic parameters of the host star. Here the default aperture size used for extraction of {\tt K2PHOT} light curve \citep{Petigura2015, Aigrain2016} is supplied to constrain the maximum allowed separation between the target and the source of the transit event. A search for the secondary eclipse was also undertaken using the method described in \citet{Dressing2017}, between values of 0.3-0.8 in orbital phase. It was found that the existing data could exclude events deeper than $5\times10^{-5}$ in units of the normalized flux. This value is then used by {\tt VESPA} as a limit on the allowable secondary eclipse. High resolution contrast curves can also be supplied to constrain the authenticity of transit signal. The Keck $K$-band contrast curve (\S\ref{sec:imaging}) along with other publicly available contrast curves like Gemini $r$-band and Gemini $z$-band\footnote{Also retrieved from ExoFOP, uploaded by David Ciardi} acquired using NIRI \citep{Hodapp2003} are supplied as input. These contrast curves are especially useful for the validation process as they can preclude the presence of nearby or background stars above a certain brightness at a given distance on sky.

\begin{figure*}[!ht]
	\centering
	\includegraphics[width=0.8\textwidth]{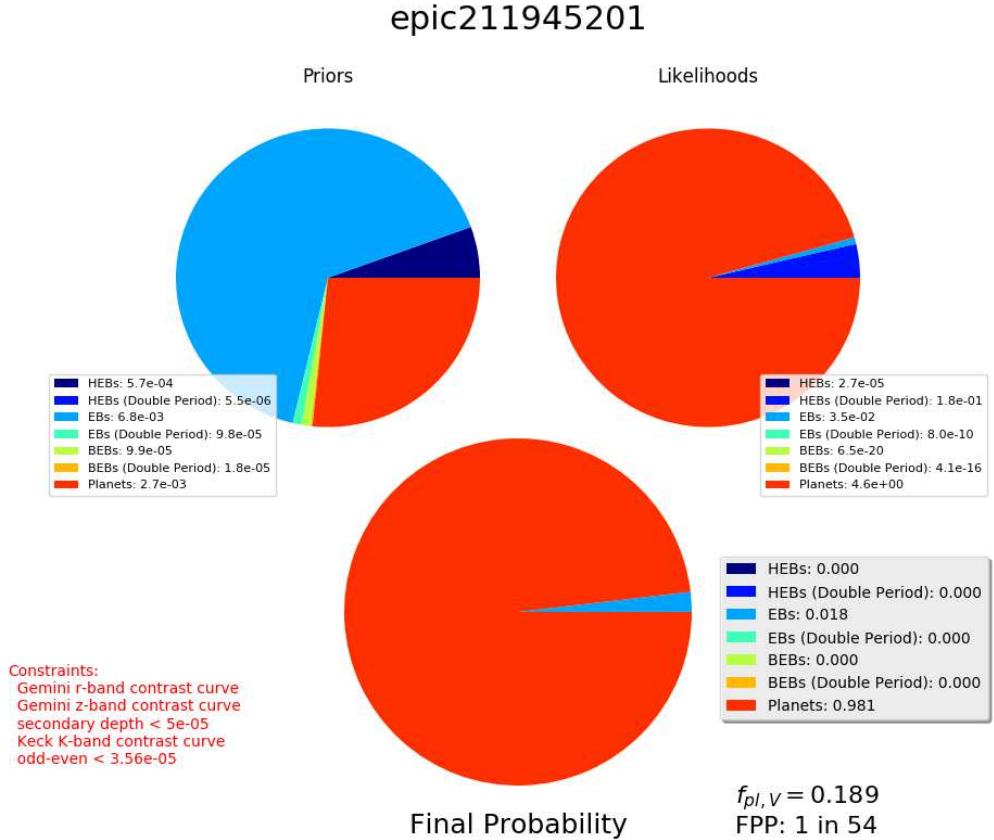}
\caption{False Positive Probability (FPP) analysis using the {\tt VESPA} package from \citet{Morton2012,Morton2015}. The incorporates the prior likelihood of a false positive scenario (given sky position, AO-imaging contrast curve data, and wavelength-dependent magnitudes), with the likelihood of the transit photometry under various scenarios. The results shown here are obtained by using the combined light curve of EPIC~211945201 from Campaigns 5 \& 16. \label{fig:Vespa}}
\end{figure*}


The properties of the host star from ExoFOP \citep[e.g. coordinates, magnitudes in various bands, see][]{Huber2016}  and other parameters summarized in Table~\ref{tab:Vespa_constraints}, were also supplied as input. The values of spectroscopic properties (mainly temperature, surface gravity and metallicity) used for the FPP calculations were obtained through spectral analysis of data obtained with the PARAS spectrograph and are listed in Table~\ref{tab:paras_spec} and further discussed in Section~\ref{sec:isochrones}.

We performed the {\tt VESPA} analysis separately for Campaign 5 \& Campaign 16 data and found the probability for the planetary scenario to be $\sim$87\% \& $\sim$96\%, and the false positive probability (FPP) to be $\sim$13\% \& $\sim$4\% respectively. We note that much of this work was initially motivated by the individual results of Campaign 5, which were much less supportive of the planetary interpretation. Ultimately we used the combined light curve from both the campaigns to calculate a probability of $\sim$98\% (Fig~\ref{fig:Vespa}) for the planetary scenario, with the remaining $\sim$2\% probability attributed to an eclipsing binary (EB) scenario. The threshold for planetary validation is a stringent $<1$\% FPP \citep{Crossfield2016}, which relegated EPIC~211945201 to the status of planetary candidate, despite the coverage of multiple transit events. Since the results of the FPP calculation from {\tt VESPA} could not single-handedly rule out the possibility of an EB, follow up with high-precision Doppler spectroscopy was necessary to establish the planetary nature of the candidate.


\begin{deluxetable}{lcc}
\small
\tabletypesize{\scriptsize}
\tablecaption{Stellar parameters supplied for Fpp calulations \label{tab:Vespa_constraints}}
\tablehead{
\colhead{Parameters} & \colhead{Value} &\colhead{source} \vspace*{15pt}\\
}
\startdata
\textbf{Main identifiers}\\
$\alpha$(J2000)(hh:mm:ss) 	&	09:06:17.75	& EPIC \\
$\delta$(J2000)(Degrees) 	&	19:24:08.11 & EPIC\\
2MASS 					    &	J09061775 + 1924080 & EPIC\\
EPIC						&	211945201 & EPIC\\
TYC							&   1404-1186-1 & EPIC\\
\hline
\textbf{Magnitudes}\\
B                 			&	10.937$\pm0.080$ & EPIC\\
g                 			&	10.479$\pm0.030$ & EPIC\\
V                 			&	10.154$\pm0.056$ & EPIC\\
r                 			&	10.038$\pm0.040$ & EPIC\\
Kep               			&	10.115 & EPIC\\
i                 			&	9.959$\pm0.080$ & EPIC\\
J                 			&	9.144$\pm0.023$ & 2MASS\\
H                 			&	8.908$\pm0.028$ & 2MASS\\
K                 			&	8.837$\pm0.020$ & 2MASS\\
W1                 			&	8.822$\pm0.023$ & WISE\\
W2                 			&	8.844$\pm0.020$ & WISE\\
W3                 			&	8.810$\pm0.028$ & WISE\\
W4                 			&	8.623$\pm0.398$ & WISE\\
\hline
\enddata
\tablecomments{Parameters with source flagged as EPIC are taken from the Ecliptic Plane Input Catalogue available at \url{http://archive.stsci.edu/k2/epic/search.php.} Other parameters whose sources are flagged as 2MASS and WISE are taken from \citep{Cutri2003} and \citep{Cutri2013}, respectively.}
\end{deluxetable}

\section{Spectroscopic Follow-Up Observations}
\label{sec:rv_obs}

In order to confirm the planetary nature of the candidate, high-resolution ($R\sim67,000$) spectroscopic follow-up observations were undertaken with the PARAS spectrograph \citep{Chakraborty2014} mounted on a 1.2 meter telescope at Gurushikhar Observatory, Mount Abu, India. PARAS is a fiber-fed, temperature and pressure stabilized, white pupil echelle spectrometer that has earlier been shown to achieve $\sim1$ m s$^{-1}$ RV precision on timescales of a month \citep{Roy:2016}. A total of 19 spectra were acquired between 25th November 2016 and 18th January 2018 using the simultaneous wavelength calibration mode (using a ThAr hollow cathode lamp) as explained in \citet{Chakraborty2014}. Besides science exposures, five bias frames and three flat frames were also acquired on each night in order to correct the bias \& verify the cross-dispersed order locations on the stabilized instrument. After each science exposure ThAr-ThAr exposures were also acquired, illuminating both the science and calibration fiber with the ThAr lamp, to correct the instrumental as well as inter-fiber drift. More details of the spectrograph, observational procedure, and data analysis techniques are described in \citet{Chakraborty2014}. The source was observed only in dark (moon-less) and photometric sky conditions. The airmass for all the epochs were $<$1.5 and seeing was better than 1.5".  The exposure time for each observation was between 1800s$-$3000s as the source is towards the fainter limit of PARAS (see Table~\ref{tab:rv_table}) which resulted in signal-to-noise ratios (SNR) between 13 to 20 per pixel at the blaze peak wavelength of 5500~\AA. A list of epochs and observational details is shown in Table~\ref{tab:rv_table}. The first column  in the table represents observation time stamp in terms of BJD-TDB. The observed RV values are listed in the next column followed by the RV errors, which are based on both photon noise and the fitting errors of the cross correlation function \citep[CCF; for CCF error estimations see][]{Chaturvedi2016a}. The RV data spans more than a year -- nearly 420 days.

\subsection{RV analysis}
\label{sec:rv_analysis}

The entire data reduction and RV analysis for PARAS was carried out by the automated pipeline in {\tt{IDL}} based on the {\tt{REDUCE}} optimal extraction routines of \cite{Piskunov2002} and explained in \citet{Chakraborty2014}. In order to reduce the data, the pipeline performs routine tasks like bias subtraction, order trace verification, cosmic ray correction (especially important due to the deep depletion CCD in PARAS), and optimal extraction of both target and calibration spectra. Instrumental drifts are tracked and corrected by cross-correlation of the simultaneous calibration spectrum with a custom ThAr mask made for the PARAS lamp. RVs are derived by cross-correlating the target spectra with a suitable numerical stellar template mask. The stellar mask is created from a synthetic spectrum of the star, containing the majority of deep photospheric absorption lines. See \cite{Baranne1996}, \cite{Pepe2002} and references therein for a more detailed description of the mask cross-correlation method. Based on the temperature given in  Table~\ref{tab:paras_spec}, which is determined from spectral analysis (see Sec~\ref{sec:isochrones}), our source is found to be a F9/G0 spectral type star. Thus, we use a G2-type stellar mask for cross-correlation of the spectra. RV measurement errors, which are given in Table~\ref{tab:rv_table}, are based on photon noise errors \citep{Bouchy2001} and the errors associated with the CCF fitting function. These errors range from 9 to 16 m~s$^{-1}$; the method for their computation is described in \cite{Chaturvedi2016a}.

\begin{figure*}[!ht]
	\includegraphics[width=0.5\textwidth]{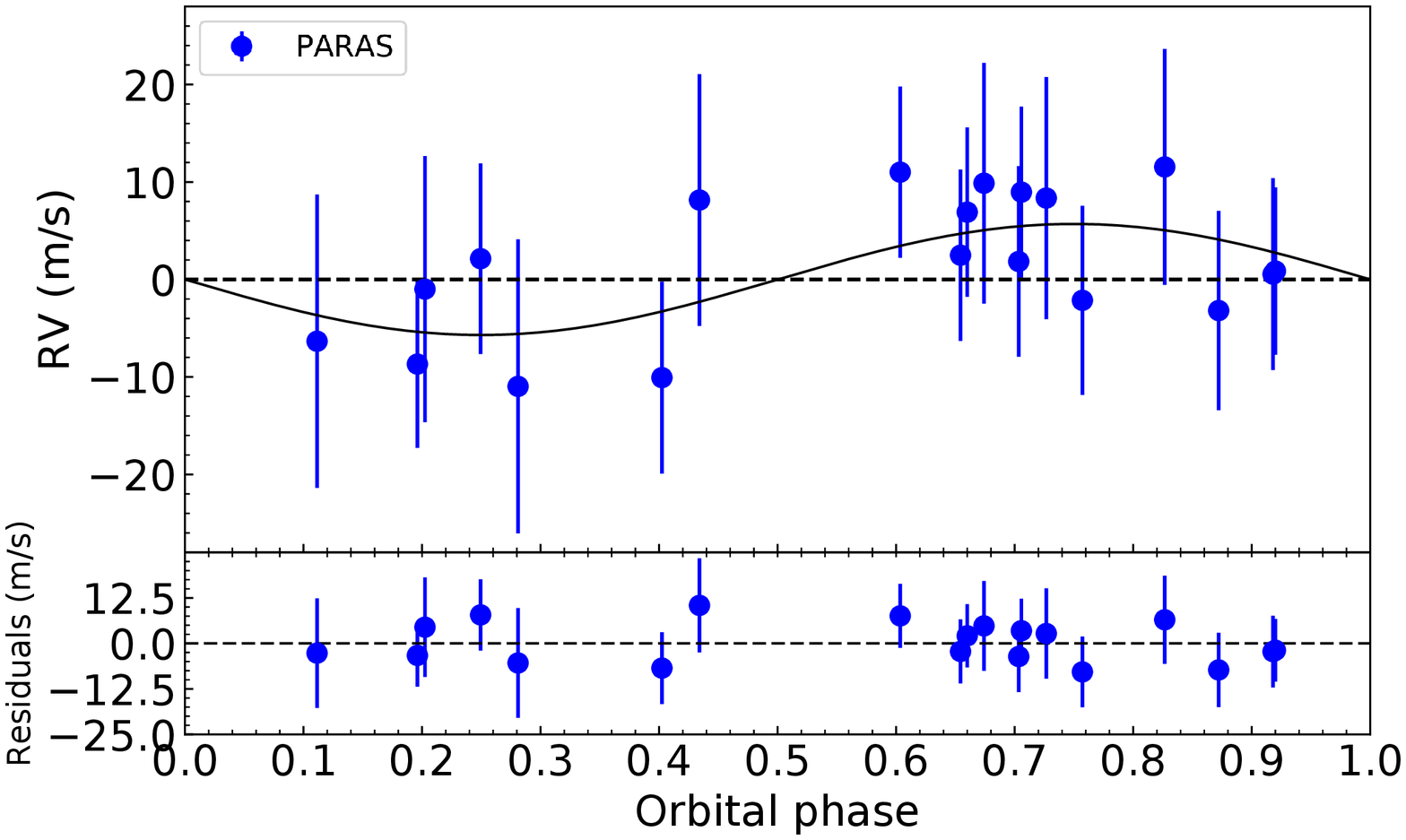}
	\includegraphics[width=0.5\textwidth]{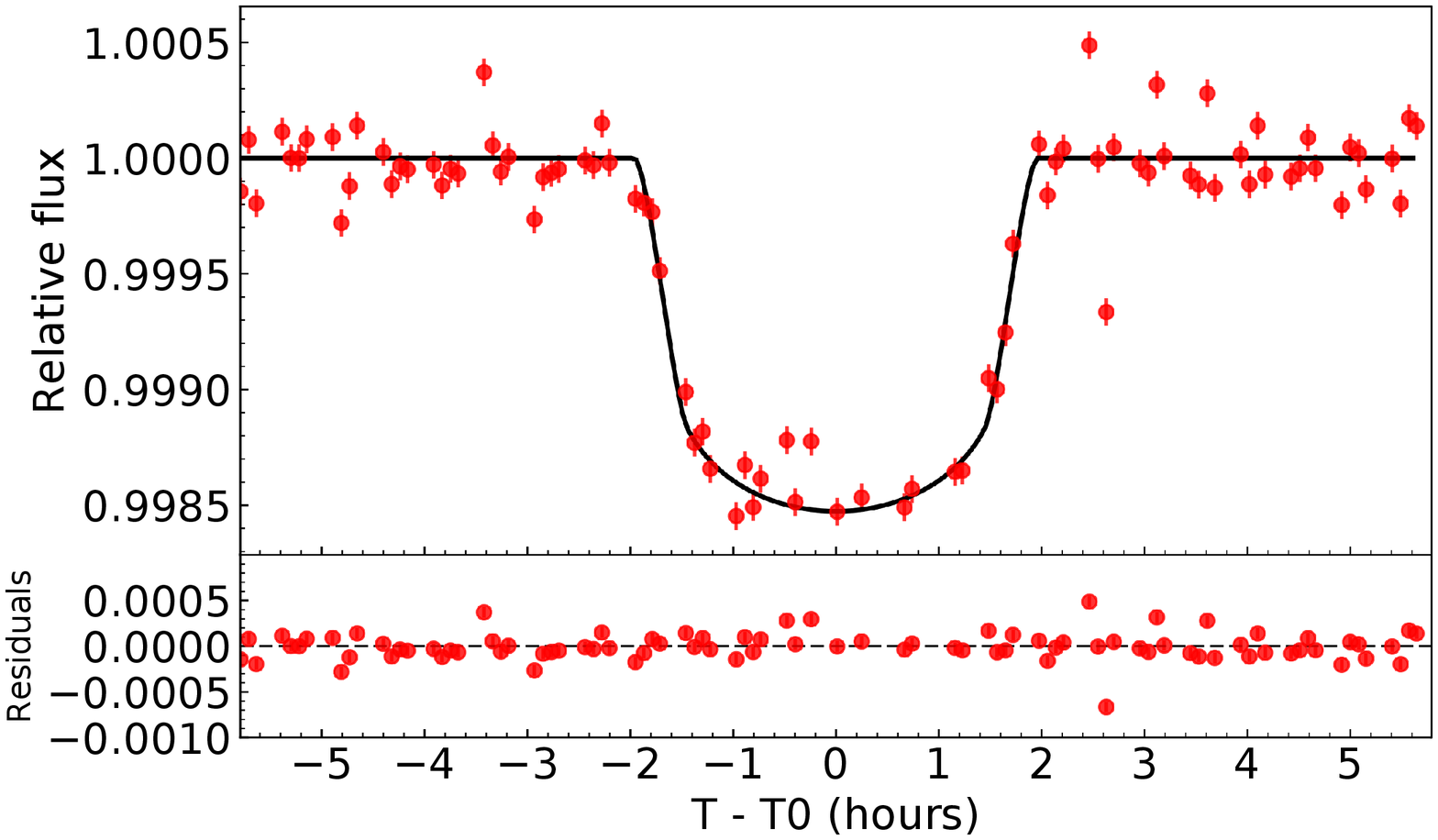}
\caption{Left: RVs taken by PARAS, phased to the 19.491d period as determined from the simultaneous fitting of the photometric and RV data using {\tt{PYANETI}}. The best fit model is also displayed together with the residuals. Right: Phase folded {\tt K2PHOT} light curve centered on the $T_0$ with the best fit model in black overlaid.\label{fig:rv_plot}}
\end{figure*}

\begin{figure*}[!ht]
\centering
	\includegraphics[width=0.6\textwidth]{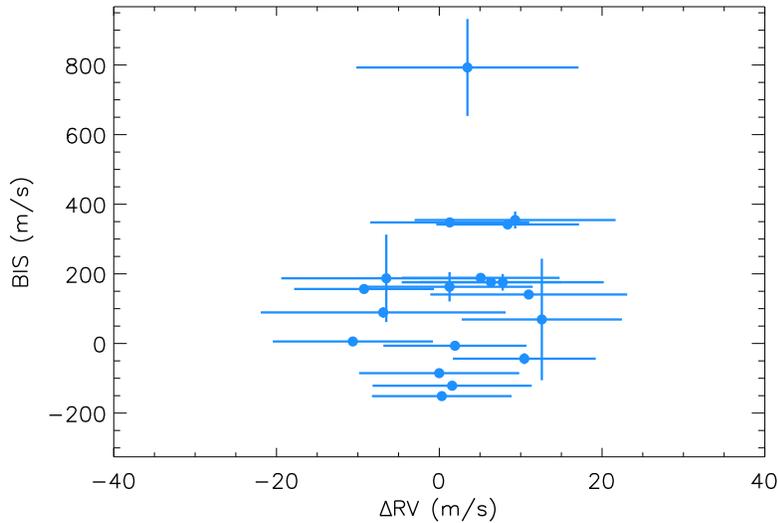}
	\caption{Distribution of measured radial velocities from PARAS, and associated values of the bisector inverse slope from the CCF. There is no significant correlation between the two, ruling out the possibility of blended spectra.
	\label{fig:bisector_plot}}
\end{figure*}

The radial  velocities from PARAS are plotted in Fig. ~\ref{fig:rv_plot}, along with the best fit model (determined by the joint fit with the photometric data) described in Sec~\ref{sec:pyaneti}. In order to check radial velocity variation induced by a blended spectrum, we computed the bisector inverse slope (BIS) of the cross-correlation function for each observation in the manner of \citet{Queloz2001}. The value of the BIS for each CCF, and its respective errors, are listed in Table~\ref{tab:rv_table} and plotted in Fig ~\ref{fig:bisector_plot}. We find no correlation between the bisector inverse slope and the measured radial velocity. If the signal detected was due to a blended spectrum, then we would expect to see a strong correlation between the bisector slopes and radial velocity measurements \citep[e.g.][]{Wright:2013}.

\subsection{Discarding an EB scenerio}
\label{sec:central_part}

\begin{figure*}[!ht]
\centering
	\includegraphics[width=1.0\textwidth, trim=0 1cm 0 3.5cm, clip=True]{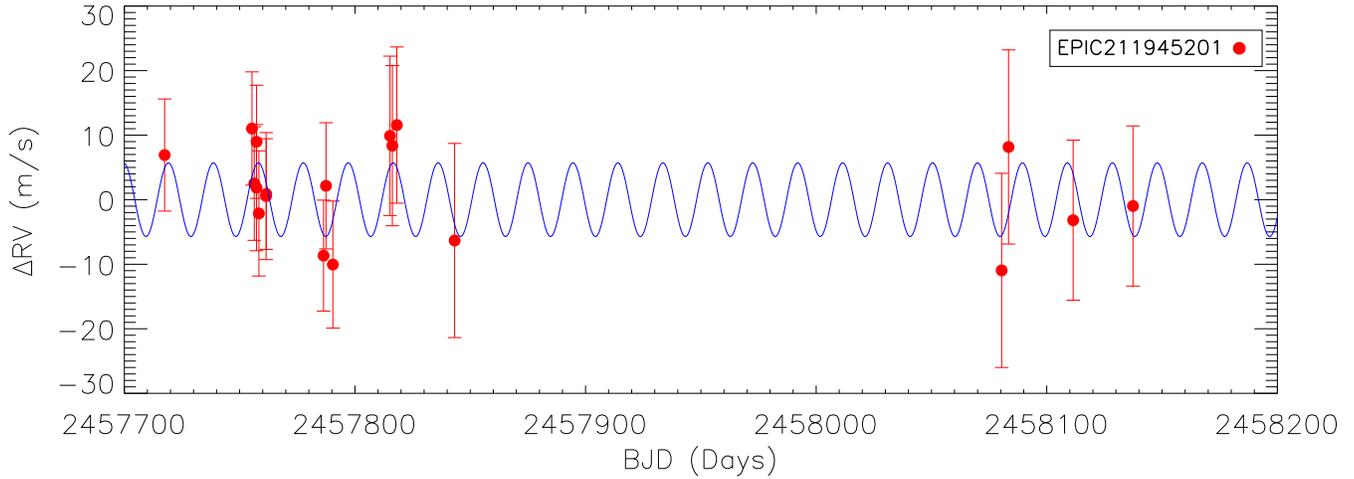}
\caption{Radial velocities for EPIC~211945201 over a span of 420 days, observed with the PARAS spectrograph. The absence of large RV dispersion is clearly seen, which discards an EB scenario(see Sec~\ref{sec:central_part}). The overlaid blue curve is the best fit model to our dataset (from PYANETI). \label{fig:rv_all}}
\end{figure*}

The radial velocities observed with PARAS are listed in Table~\ref{tab:rv_table} and plotted in Fig~\ref{fig:rv_all}. The change in the RV values throughout our observational span is small -- within $\pm20$ m~s$^{-1}$ -- which discards the possibility that this is an EB system. The RV dataset combined with FPP results (Sec~\ref{sec:vespa}) shows very strong evidence that the body revolving around EPIC~211945201 in a 19.491 day orbit is indeed a planet.

\startlongtable
\begin{deluxetable*}{lcccccc}
\tabletypesize{\scriptsize}
\tablecaption{Radial Velocities of EPIC~211945201 in chronological order \label{tab:rv_table}}
\tablehead{
\\\colhead{BJD$_{TDB}$} & \colhead{RV} & \colhead{$\sigma$-RV} & \colhead{BIS} & \colhead{$\sigma$-BIS} & \colhead{Exp. Time} \\
\\\colhead{} & \colhead{(km s$^{-1}$)} & \colhead{(km s$^{-1}$)} & \colhead{(km s$^{-1}$)} & \colhead{(km s$^{-1}$)} & \colhead{(sec.)}\\
}
\startdata
2457717.469599   &  1.2657   &   0.0087  &      0.1763  &      0.0105      & 3000 \\
2457755.352017   &  1.2698   &   0.0088  &   $-$0.0437  &    0.0155	& 3000 \\
2457756.343921   &  1.2613   &   0.0089  &   $-$0.0066  &    0.0110	& 3000 \\
2457757.300892   &  1.2607   &   0.0098  &      0.3477  &      0.0102	& 3000 \\
2457757.344621   &  1.2678   &   0.0088  &      0.3419  &      0.0132	& 3000 \\
2457758.348027   &  1.2567   &   0.0097  &      0.1885  &      0.0099	& 3000 \\
2457761.483625   &  1.2594   &   0.0098  &   $-$0.0853  &    0.0112	& 3000 \\
2457761.522539   &  1.2597   &   0.0086  &   $-$0.1513  &    0.0107	& 3000 \\
2457786.395919   &  1.2501   &   0.0086  &      0.1565  &      0.0080	& 2400 \\
2457787.435365   &  1.2610   &   0.0098  &   $-$0.1214  &    0.0086	& 2400 \\
2457790.417556   &  1.2487   &   0.0098  &      0.0058  &      0.0099	& 2400 \\
2457815.206300   &  1.2687   &   0.0123  &      0.3546  &      0.0244	& 1800 \\
2457816.230661   &  1.2671   &   0.0124  &      0.1758  &      0.0239	& 1800 \\
2457818.178771   &  1.2704   &   0.0121  &      0.1407  &      0.0127	& 1800 \\
2457843.222090   &  1.2525   &   0.0150  &      0.0891  &      0.0148	& 2400 \\
2458080.433400   &  1.2528   &   0.0151  &      0.1871  &      0.1255	& 2400 \\
2458083.416314   &  1.2720   &   0.0129  &      0.0690  &      0.1746	& 2400 \\
2458111.447097   &  1.2606   &   0.0102  &      0.1629  &      0.0422 	& 2400 \\
2458137.379171	 &  1.2628   &   0.0136  &      0.7929  &      0.1392	& 2400 \\
\hline
\enddata
\end{deluxetable*}

\begin{figure*}[!ht]
\centering
	\includegraphics[width=1.0\textwidth, trim=0 1cm 0 3.5cm, clip=True]{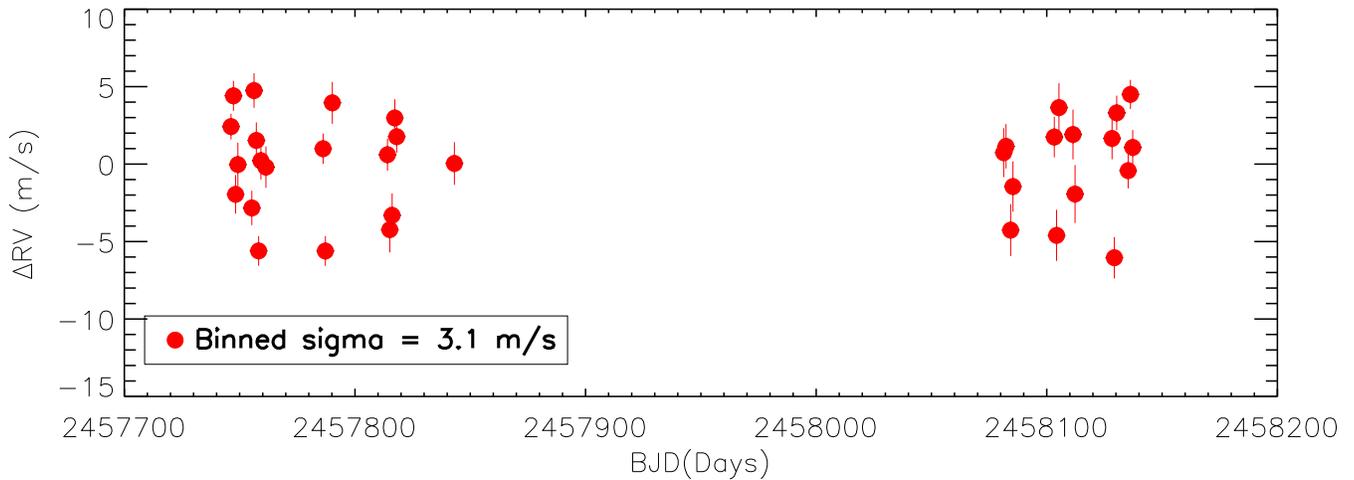}
\caption{Radial velocities for standard star HD55575. The RV points in red show nightly binned averages. The data spans over 391 days and demonstrates instrument stability and performance during the acquisition of the EPIC~211945201 data.\label{fig:HD55575}}
\end{figure*}

To verify the instrument stability of the PARAS spectrograph over the same period of time, the RV standard star HD55575 \citep{Bouchy2013} was monitored closely alongside EPIC~211945201. This star was observed by {\tt{SOPHIE}} \citep{Bouchy2013, Perruchot2008} in HR+ mode for a span of 118 days. The RV dispersion (${\sigma_{RV}}$) of this target was reported to be 3.4 m~s$^{-1}$ in \citet{Bouchy2013}. We acquired 58 spectra of HD55575 over a span of 391 days. The RVs were then calculated identical to the manner described in Sec~\ref{sec:rv_analysis} for EPIC~211945201. The ${\sigma_{RV}}$ with PARAS was found to be 3.5 m~s$^{-1}$. Nightly binning of the data points reduced this dispersion further to 3.1 m~s$^{-1}$. This demonstrates that our RV dispersion is consistent with that achieved by {\tt{SOPHIE}} for HD55575, and quantifies the upper limit of the spectrograph's stability during the entire long span of observations of EPIC~211945201. With strong evidence that our source hosts an exoplanet, we proceed to estimating the mass of the transiting body by simultaneous modeling of RV and photometry data in the Sec~\ref{sec:pyaneti}.

\section{Physical parameters of the star}
\label{sec:isochrones}

The spectral parameters of the star were first estimated using the {\tt{PARAS SPEC}} package \citep{Chaturvedi2016a}. {\tt{PARAS SPEC}} is a stellar synthesis pipeline that estimates $T\rm{_{eff}}$, $\log(g)$, $[Fe/H]$, $v sin i$ $\&$ $v_{\rm{micro}}$ by implementing both synthetic spectral fitting and equivalent width measurements. The details of the package can be found in \citet{Chaturvedi2016a} \& \citet{Chaturvedi2016b}. For EPIC~211945201, the coadded high SNR spectra for all the epochs listed in Table~\ref{tab:rv_table} were used for this analysis. The adopted values for stellar parameters are the weighted average of the results obtained from both the methods of {\tt{PARAS SPEC}}. Stellar parameters present in the literature, along with our new estimates, are listed in Table ~\ref{tab:paras_spec}. It can be seen that our estimated parameters are within the 2-$\sigma$ confidence interval of the parameters available in the literature. These estimated stellar parameters, along with the photometric magnitudes given in Table~\ref{tab:Vespa_constraints} of the star in different bands, are used to derive the physical parameters of the star.
The new precision astrometry from the GAIA DR2 data release \citep{GaiaCollaboration2018} is also used in this derivation of stellar parameters. The parallax for the star measured by GAIA is 5.475$\pm0.039$\footnote{https://gea.esac.esa.int/archive/}. The Dartmouth Stellar evolution database \citep{Dotter2008} is utilized to obtain the radius, mass, age and distance to the host star using the {\tt{ISOCHRONE}} package \citep{Morton2015}. The uncertainties associated with photometric and spectroscopic observables were also taken into account while estimating the properties of the star. Finally, a radius of $R_{*}$ = $1.38_{-0.018}^{+0.017}$ $R_{\odot}$, mass of $M_{*}$ = $1.18_{-0.04}^{+0.03}$ $M_{\odot}$, age of $3.99_{-0.7}^{+0.85}$ Gyr, and a distance of 182.6$\pm1.3$~pc are determined for the star. The very precise parallax measurements from GAIA lead to an extremely precise determination of stellar parameters, shrinking the error bars by an order of magnitude, and exhibiting the tremendous value of this mission.

\begin{deluxetable}{lccc}
\small
\tabletypesize{\scriptsize}
\tablecaption{Stellar parameters\label{tab:paras_spec}}
\tablehead{
\colhead{Parameters} &\colhead{\citep{Petigura2018}} &\colhead{\citep{Mayo2018}} &\colhead{This work} \vspace*{10pt}\\
}
\startdata
$\teff$[K]	& 6018$\pm60$ & 6046$\pm50$ &6025$\pm100$\\
$\log(g)$[dex] & 4.13$\pm0.1$ &4.14$\pm0.1$ &4.25$\pm0.1$ \\
$\feh$[dex] & 0.12$\pm0.04$ &0.05$\pm0.08$ &0.1$\pm0.1$\\
$v_{\rm{micro}}$ [Km/s] & - & - & 0.4$\pm0.1$ \\
$v_{\rm{rot}}$  [Km/s] &3.3$\pm1.0$ & - & 4.0$\pm1.0$\\
\hline
\enddata
\end{deluxetable}

\section{Simultaneous Fitting and modeling of RV and Photometry data}
\label{sec:pyaneti}

The {\it K2} photometry data and PARAS radial velocity data are simultaneously fitted using the {\tt{PYANETI}} routine of \citep{Barragan2016,Barragan2017}. This {\tt{PYTHON}} code uses Markov chain Monte Carlo (MCMC) methods with a Bayesian approach and a parallelized ensemble sampler algorithm in {\tt Fortran}. The photometric dataset included in the joint analysis was subset of the whole {\tt k2sff} light curve. About 13 hours of data centered on each of the seven transits observed by {\it K2} is selected. The final dataset contains 194 photometric data points and 19 RV data points as listed in Table~\ref{tab:rv_table}.

The datasets were fitted assuming a circular Keplerian orbit, i.e. we fixed eccentricity $e=0$ and longitude of periastron $\omega=90^{\circ}$. Other orbital parameters were allowed to float, including the systemic velocity $\gamma$ for the PARAS instrument, the RV semi-amplitude $K$, mid-transit time $T_{0}$, orbital period $P_{orb}$, impact parameter $b$, semi major axis in terms of stellar radius $a/R_{*}$,$q1$, $q2$, and the planet to star radius ratio $R_{p}/R_{*}$. Here, $q1$ \& $q2$ are the parameterization of $u_{1}$ and $u_{2}$ as described in \citet{Kipping2003}. Also, in order to fit the photometry data the quadratic limb-darkening law of \citet{Mandelagol2002} was followed.

We generated 250,000 independent points for each free parameter by exploration of parameter space using 500 Markov chains. In order to find the global solution for the dataset, a wide range of uniform priors $P_{orb}$ $=$ $[19.40,19.60]$ days, $T_{0}$ $=$ $[2458112.5403,2458114.6532]$, $b$ $=$ $[0,1]$, $a/R_{*}$ $=$ $[5,100]$, $R_{p}/R_{*}$ $=$ $[0.005,0.1]$, $K$ $=$ $[0.001,1.0] $ km~s$^{-1}$, $\gamma_{j}$ $=$ $[1,100] $ km~s$^{-1}$ and $q_{1}$,$q_{2}$ $=$ $[0,1]$ were chosen. The final derived parameter values and their associated uncertainties are given by the median and 68.3$\%$ confidence interval of the posterior probability distribution; these are listed in Table~\ref{tab:result_final}. The global fitting of the limb darkening coefficients gives $q_1$ = $0.149_{-0.075}^{+0.077}$ \&  $q_2$  = $0.457_{-0.308}^{+0.384}$, which results in $u_1$ = $0.327_{-0.231}^{+0.345}$ \& $u_2$ = $0.028_{-0.289}^{+0.220}$. As the uncertainties associated with these are very high, we instead interpolate the limb darkening coefficients given in \citet{Claret2011} for the Kepler-band, and also list these values in Table~\ref{tab:result_final}.

\startlongtable
\begin{deluxetable*}{lcl}
\tabletypesize{\scriptsize}
\tablecaption{MEDIAN VALUES AND 68 \% CONFIDENCE INTERVALS FOR THE LIGHCURVE AND RADIAL VELOCITY PARAMETERS OF THE EPIC~211945201 SYSTEM\label{tab:result_final}}
\tablehead{\colhead{~~~Parameter} & \colhead{Units} & \colhead{Value} \vspace*{20pt}}
\startdata
\sidehead{Stellar Parameters:}
                           ~~~$M_{*}$\dotfill &Mass (\msun)\dotfill & $1.18_{-0.04}^{+0.03}$\\
                        ~~~$R_{*}$\dotfill &Radius (\rsun)\dotfill & $1.38_{-0.018}^{+0.017}$\\
                         ~~~$\rho_*$\dotfill &Density (cgs)\dotfill & $0.63_{-0.03}^{+0.04}$\\
             ~~~$\log(g_*)$\dotfill &Surface gravity (cgs)\dotfill & 4.25$\pm0.1$\\
              ~~~$\teff$\dotfill &Effective temperature (K)\dotfill & 6025$\pm100$\\
                              ~~~$\feh$\dotfill &Metallicity\dotfill & 0.1$\pm0.1$\\
				      ~~~$Age$\dotfill &Gyr\dotfill & $3.99_{-0.7}^{+0.85}$\\
				 ~~~$Distance$\dotfill &pc\dotfill  & 182.6$\pm1.3$\\
       ~~~$v_{\rm{micro}}$\dotfill &Microturbulence (Km/s)\dotfill & 0.4$\pm0.1$\\
      ~~~$v_{\rm{rot}}$\dotfill &Rotational Velocity (Km/s)\dotfill & 4.0$\pm1.0$\\
\sidehead{Planetary Parameters:Fitted}
                ~~~$T_0$\dotfill &Time of transit (\bjdtdb)\dotfill & $2458113.93994_{-0.00039}^{+0.00039}$\\
                               ~~~$e$\dotfill &Eccentricity\dotfill & $0.0$(fixed)\\
    ~~~$\omega_*$\dotfill &Argument of periastron (degrees)\dotfill & $90.0$(fixed)\\
                              ~~~$P$\dotfill &Period (days)\dotfill & $19.49213_{-0.00001}^{+0.00001}$\\
                           ~~~$b$\dotfill &Impact Parameter\dotfill & $0.85_{-0.008}^{+0.007}$\\
     ~~~$a/R_{*}$\dotfill &Semi-major axis in stellar radii\dotfill & $23.10_{-0.47}^{+0.47}$\\
~~~$R_{P}/R_{*}$\dotfill &Radius of planet in stellar radii\dotfill & $0.0407_{-0.0003}^{+0.0003}$\\
		    ~~~$K$\dotfill &RV semi-amplitude (m/s)\dotfill & $5.7_{-2.7}^{+3.0}$\\
\sidehead{Planetary Parameters:Derived}
                      ~~~$i$\dotfill &Inclination (degrees)\dotfill & 87.90$\pm0.06$\\
                       ~~~$a$\dotfill &Semi-major axis (AU)\dotfill & 0.148$\pm0.004$\\
                    ~~~$M_{P}$\dotfill &Mass ($M_{\oplus}$)\dotfill & $27_{-12.6}^{+14}$\\
                  ~~~$R_{P}$\dotfill &Radius ($R_{\oplus}$)\dotfill & 6.12$\pm0.10$\\
                       ~~~$\rho_{P}$\dotfill &Density (cgs)\dotfill & $0.65_{-0.30}^{+0.34}$ \\
                          ~~~$g_{P}$\dotfill & gravity(cgs)\dotfill & $689.96_{-322.20}^{+360.87}$\\
            ~~$T_{P}$\dotfill &Time of periastron (\bjdtdb)\dotfill & $2458113.93994_{-0.00039}^{+0.00039}$\\
           ~~~$T_{eq}$\dotfill &Equilibrium Temperature (K)\dotfill & $886.35_{-17.10}^{+17.36}$\\
        ~~~$T_{23}$\dotfill &Total Eclipse duration (hours)\dotfill & $2.92_{-0.04}^{+0.04}$ \\
	        ~~~$T_{14}$\dotfill &Total duration (hours)\dotfill & 3.91$\pm0.02$\\
\sidehead{Other Parameters:}
              ~~~$u_1$\dotfill &linear limb-darkening coeff\dotfill & $0.36$\tablenotemark{a}\\
           ~~~$u_2$\dotfill &quadratic limb-darkening coeff\dotfill & $0.29$\tablenotemark{a}\\
                       ~~~$q_1$\dotfill & \((u_1 + u_2)^2\)\dotfill & $0.42$\tablenotemark{a} \\
	    ~~~$q_2$\dotfill &  \(u_1(2(u_1 + u_2))^{-1}\)\dotfill  & $0.28$\tablenotemark{a}\\
      ~~~$\gamma_{PARAS}$\dotfill &Systemic velocity (Km/s)\dotfill & 1.2588$\pm0.0025$\\
\enddata
\tablenotetext{a}{Limb-Darkening coefficients obtained by interpolating the table of \citet{Claret2011}.}
\end{deluxetable*}



\section{Disscussion}
\label{sec:internal_comp}

The simultaneous fitting of RV and photometric data gives a RV semi-amplitude of $K=5.7_{-2.7}^{+3.0}$ m~s$^{-1}$, and mass of  $M_p = 27_{-12.6}^{+14}$ $M_{\oplus}$ for EPIC~211945201b. This gives a 3-$\sigma$ upper limit on the mass of 42~$M_{\oplus}$ above the nominal value.
Leveraging the precision of the newly released GAIA DR2 data, the radius of the planet is derived to be 6.12$\pm0.10$ $R_{\oplus}$. This radius and mass correspond to a density of $0.65_{-0.30}^{+0.34}$~g~cm$^{-3}$. It is be noted here that earlier radius estimates of $6.0^{+0.9}_{-0.8}$~$R_{\oplus}$ by \citet{Petigura2018} and $5.85^{+0.95}_{-0.78}$~$R_{\oplus}$ by \citet{Mayo2018} from Campaign 5 transit data alone; and $5.3$~$R_{\oplus}$ by \citet{Yu2018} from Campaign 16 data, are in agreement with our simultaneous fitting results. The planet thus has a Saturn-like density while its mass and radius place it in the super-Neptune or sub-Saturn classification of exoplanets. This class of planets is currently very underpopulated \citep{Bonomo2014}.

EPIC~211945201b lies in the overlapping region between super-Neptunes and sub-Saturns. There are only 22 sub-Saturns in the range of 4-8~$R_{\earth}$ whose mass and radius are precisely known, as discussed and listed in Table 7 of \citet{Petigura2017}. In spite of the small range in radii, these planets are found to have a large mass range, and thus densities that vary between 2.40 to 0.26 g~cm$^{-3}$, indicating diverse formation mechanisms \citep{Bayliss2015}. There is no obvious correlation between the radius and mass of these planets, making them an interesting class of exoplanets meriting further scrutiny.

The total transit duration time $T_{14,central}$, if the planet is passing in front of the stellar disk center, should be $\sim$6 hours using the equations of \citet{Seager2003}. However, fitting the light curve (Sec~\ref{sec:pyaneti}) gives $T_{14}$ = 3.91 hours, and an impact parameter of $b$ = 0.85, suggesting that the planet is passing closer to the poles of the stellar disk. This indication of a grazing transit scenario likely results in the $\sim$2 \% probability of an eclipsing binary in the {\tt VESPA} output. However, the RV observations from PARAS conclusively rule out the EB scenario and confirm the planetary nature of the source.

\begin{figure*}[!ht]
\centering
	\includegraphics[width=0.6\textwidth]{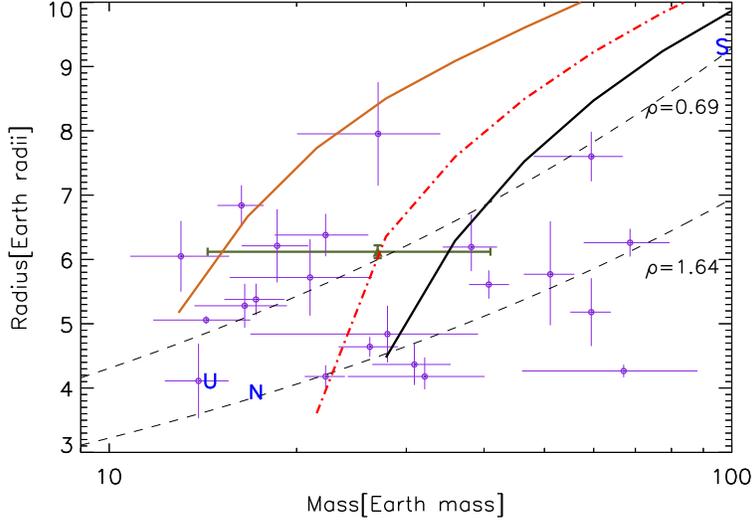}
	\caption{Mass versus radius diagram showing our measurements for EPIC~211945201b (olive triangle). Overlaid are the theoretical curves for 3.16~Gyr and core masses of 10 and 25 M$_\oplus$ at 0.1 AU from \citet{Fortney2007} in dark yellow and black, respectively. The dashed-dotted Red line is linearly interpolated fortney model for core of 18 M$_\oplus$(see  Sec~\ref{sec:internal_comp}). The dashed black lines from top to bottom are isodensity curves for saturn \& neptune density. The magenta coloured open circles represents the 22 superneptunes (4.0$\leq$$R_{\oplus}$$\geq$8.0, 10.0$\leq$$M_{\oplus}$$\geq$70.0) whose mass and radius are determined with at least 50 \% precision (from \url{http://exoplanet.eu/} , \citealp{Schneider2011} \& NASA exoplanet Archive, \citealp{Akeson2013}). The Blue coloured U, N, and S represents here Uranus, Neptune and Saturn, respectively.
	\label{fig:internal_comp}}
\end{figure*}

We also estimate the internal composition of EPIC~211945201b using irradiated planetary composition models of \citet{Fortney2007} \& \citet{Lopez&Fortney2014}. These are two layer partitioning models based on the assumption that planets are made up of a very dense core and a less dense envelope, and that all the heavy elements are concentrated in the core while the lighter elements are part of envelope. In Fig~\ref{fig:internal_comp} we plot our results, along with the synthetic models used to estimate the core and envelope masses. The models chosen here are based on our final parameters for the EPIC~211945201 system -- at an age of 3.16~Gyr and 0.1AU separation, with core masses of 0,10, and 25~$M_{\earth}$ -- where we interpolate between 10 and 25~$M_{\earth}$ core mass models. We determine the heavy element content to be about 60-70\% of the total mass. The non-irradiated models of \citet{mordasini2012} \& \citet{jin&mordasini2014} were also used to find the heavy element fraction and the results found to be consistent with 60-70\%. However, we caution that these values have high uncertainties given our errors on the planet density, and should be considered preliminary estimates until better mass measurements are obtained.


\section{Summary}
\label{sec:summary}
We find strong evidence of a transiting sub-Saturn (or super-Neptune) around EPIC~211945201. Previous work had deemed this a planetary candidate, which we validate with {\tt VESPA} analysis. We found the FPP for the planetary hypothesis to be $>$1\%, with a $\sim$2 \% probability that the system is an EB. In order to resolve the situation, we conducted a campaign of high-resolution RV observations using the PARAS spectrograph. These RV measurements showed low dispersion over a long time baseline, eliminating the EB scenario and confirming the planetary nature of the candidate. By simultaneously modeling the RV and {\it K2}-photometry data we derive a radius and mass of 6.12$\pm0.1$~$R_{\earth}$ and $27_{-12.6}^{+14}$~$M_{\oplus}$ for the companion. We then used the models of \citet{Fortney2007} and \citet{Lopez&Fortney2014} to make a notional prediction of the core mass, finding a substantial projected heavy element fraction of 60-70 \%, although more precise mass measurements are necessary to confirm this heavy element content. EPIC~211945201b remains interesting as it straddles the transition regime between ice giants and gas giants. The detection of similar exoplanets will continue to hone our understanding of the formation and distribution of these worlds that have no analog in our Solar System, and yet seem to abound in the nearby universe.\\

\textbf{Future Works:}
\label{sec:future}
The accurate mass of EPIC~211945201b can be determined with more precise RV measurements using HARPS \citep{Pepe2003} or the newly-commissioned HPF \citep{Mahadevan2014} spectrographs. This will provide clearer insight into the internal structure of the planet. The source is also a good target to study the star planet alignment (R-M effect) as the host star is relatively bright.

\textbf{Acknowledgments:}
The PARAS spectrograph is fully funded and being supported by Physical Research Laboratory (PRL), which is part of Department of Space, Government of India. We would like to thank Director, PRL for his support. We acknowledge the help from Vishal Shah, Kapil Kumar, Kevikumar Lad, Ashirbad Nayak and Mount Abu Observatory staff for their support during observations. We acknowledge the use of SIMBAD database operated at the CDS, Strasbourg, France. We also acknowledge the support from ExoFOP users, who willingly share the follow up observational data with the community. This research has made use of the NASA Exoplanet Archive, which is operated by the California Institute of Technology, under contract with the National Aeronautics and Space Administration under the Exoplanet Exploration Program; and the Extrasolar Planets Encyclopaedia (exoplanet.eu). We are also thankful to Francesco Pepe (Geneva Observatory) for providing G2 stellar mask which is used for RV reduction.

\software{{\tt{VESPA}} (\citealp{Morton2012}, \citealp{Morton2015}),
		{\tt{PARAS PIPELINE}} \citep{Chakraborty2014},
		{\tt{PARAS SPEC}} \citep{Chaturvedi2016a,Chaturvedi2016b},
		 {\tt{ISOCHRONES}} \citep{Morton2015},
		 {\tt{PYANETI}} \citep{Barragan2016,Barragan2017}}

\bibliographystyle{aasjournal}
\bibliography{reference}
\end{document}